\documentclass[preprintnumbers,amsmath,amssymbm,preprint]{revtex4}
\usepackage{epsfig}
\usepackage{graphicx}
\usepackage{amssymb}

\begin{document}
%\title{Analytic treatment of the Kerr black-hole-mirror bomb}
%\title{Analytic treatment of the electromagnetic and gravitational black-hole bombs}
%\title{The electromagnetic and gravitational black-hole bombs}
\title{Algebraically special resonances of the
Kerr-black-hole-mirror bomb}

\author{Shahar Hod}
\affiliation{The Ruppin Academic Center, Emeq Hefer 40250, Israel}
\affiliation{ } \affiliation{The Hadassah Institute, Jerusalem
91010, Israel}
\date{\today}

\begin{abstract}
\ \ \ A co-rotating bosonic field interacting with a spinning Kerr
black hole can extract rotational energy and angular momentum from
the hole. This intriguing phenomenon is known as superradiant
scattering. As pointed out by Press and Teukolsky, the
black-hole-field system can be made unstable (explosive) by placing
a reflecting mirror around the black hole which prevents the
extracted energy from escaping to infinity. This composed
black-hole-mirror-field bomb has been studied extensively by many
researchers. It is worth noting, however, that most former studies
of the black-hole bomb phenomenon have focused on the specific case
of confined scalar (spin-$0$) fields. In the present study we
%use analytical techniques in order to
explore the physical properties of the higher-spin (electromagnetic
and gravitational) black-hole bombs. It is shown that this composed
system is amenable to an {\it analytic} treatment in the physically
interesting regime of rapidly-rotating black holes. In particular,
we prove that the composed black-hole-mirror-field bomb is
characterized by the {\it unstable} resonance frequency
$\omega=m\Omega_{\text{H}}+is\cdot 2\pi T_{\text{BH}}$ (here $s$ and
$m$ are respectively the spin-parameter and the azimuthal harmonic
index of the field, and $\Omega_{\text{H}}$ and $T_{\text{BH}}$ are
respectively the angular-velocity and the temperature of the
rapidly-spinning black hole). Our results provide evidence that the
higher-spin (electromagnetic and gravitational) black-hole-mirror
bombs are much more explosive than the extensively studied scalar
black-hole-mirror bomb. In particular, it is shown here that the
instability growth rates which characterize the higher-spin
black-hole bombs are two orders of magnitudes larger than the
instability growth rate of the scalar black-hole bomb.
\end{abstract}
\bigskip
\maketitle

%]

\section{Introduction}

Recent astrophysical observations provide compelling evidence that
rapidly-rotating Kerr black holes are ubiquitous in our Universe
\cite{Bervol,Bre,Rey,Yang3}. These spinning black holes contain
large amounts of rotational energy which can be released in various
astrophysical processes. The mass-energy $M$ of a spinning Kerr
black hole can be expressed in the form \cite{Ch,ChRu,Noteun}
\begin{equation}\label{Eq1}
M=\sqrt{{{A}\over{16\pi}}+{{4\pi J^2}\over{A}}}\  ,
\end{equation}
where $J$ and $A$ are the angular-momentum and surface area of the
black hole, respectively. Remembering that the black-hole surface
area is an irreducible quantity in classical general relativity
\cite{Haw}, one concludes that a fraction $1-1/\sqrt{2}\ (\sim
29\%)$ of the mass-energy of a maximally spinning Kerr black hole
\cite{Notemax1} is in the form of rotational energy. This is the
fraction of energy which in principle can be extracted from a
maximally spinning (extremal) Kerr black hole \cite{Noterev}.

A concrete physical mechanism to extract energy from spinning black
holes was suggested by Zel'dovich more than four decades ago
\cite{Zel}. The extraction mechanism is based on the remarkable
phenomenon of superradiant amplification: co-rotating bosonic fields
interacting with a spinning black hole \cite{NoteRN,Bekch,Hodch} can
gain energy (be amplified) on the expense of the black-hole
rotational energy (this process is also characterized by the
extraction of angular momentum from the spinning black hole).

Soon after the discovery of Zel'dovich \cite{Zel} regarding the
superradiant amplification of bosonic fields by spinning black
holes, it was pointed out by Press and Teukolsky \cite{PressTeu2}
that this remarkable mechanism can be used in order to build a
powerful {\it black-hole bomb}. This bomb is made of three
ingredients \cite{PressTeu2}: (1) a spinning black hole, (2) a
bosonic field which interacts with the black hole to extract its
rotational energy, and (3) a reflecting mirror which surrounds the
black hole and prevents the extracted energy from escaping to
infinity \cite{Notemas}.

It was suggested in \cite{PressTeu2} that, the combined effect of
superradiant amplification of the scattered bosonic field by the
spinning black hole and its subsequent confinement by the reflecting
mirror, would lead to an exponential growth in the field's amplitude
(an exponential growth in the amount of rotational energy extracted
from the central black hole). This expectation was confirmed in the
important work of Cardoso et. al. \cite{CarDias} for the specific
case of confined {\it scalar} (spin-$0$) fields. In particular, it
was established in \cite{CarDias} that, for a scalar mode of the
form $e^{im\phi-i\omega t}$ \cite{Noteher}, the critical field
frequency
\begin{equation}\label{Eq2}
\omega=m\Omega_{\text{H}}
\end{equation}
marks the onset of instability in the black-hole-mirror-scalar-field
system. (Here $\Omega_{\text{H}}$ is the angular-velocity of the
black-hole horizon. This angular velocity can be expressed in the
form $\Omega_{\text{H}}=a/2Mr_+$, where $a\equiv J/M$ and $r_+$ are
respectively the specific angular momentum and horizon-radius of the
black hole).

Specifically, it was shown in \cite{CarDias} that, for confined
scalar (spin-$0$) fields, the critical resonance frequency
(\ref{Eq2}) sits on the boundary between stable and unstable
black-hole-mirror-scalar-field configurations: confined scalar modes
with $\Re\omega>m\Omega_{\text{H}}$ decay in time (these scalar
modes are characterized by $\Im\omega<0$) whereas confined scalar
modes with $\Re\omega<m\Omega_{\text{H}}$ grow exponentially over
time (these scalar modes are characterized by $\Im\omega>0$).

In addition, using numerical techniques it was established in
\cite{CarDias} that the instability growth rate \cite{Noteimw} of a
(confined and amplified) scalar field increases with increasing
values of the black-hole spin parameter $a$. In other words, it was
shown \cite{CarDias} that rapidly-rotating Kerr black holes are
characterized by $\Im\omega$ values which are larger than the
corresponding $\Im\omega$ values of slowly-rotating black holes. In
particular, the numerical results of \cite{CarDias} reveal that the
maximum instability growth rate of the
black-hole-mirror-scalar-field bomb is characterized by
\cite{Notemax}
\begin{equation}\label{Eq3}
\Im\omega^{\text{scalar}}_{\text{max}}\simeq 6\times 10^{-5}M^{-1}\
.
\end{equation}
It is worth emphasizing that this maximal value of
$\Im\omega^{\text{scalar}}$ corresponds to confined scalar fields
which are amplified by a maximally-spinning Kerr black hole.

It is worth emphasizing that most former studies of the black-hole
bomb phenomenon have focused on the specific model of confined
scalar (spin-$0$) fields. To the best of our knowledge, no analogous
results exist in the literature for the physically interesting cases
of confined {\it higher}-spin (electromagnetic and gravitational)
fields. While the well-explored scalar black-hole-mirror bomb may
serve as a simple toy model for the energy extraction mechanism from
black holes, it should be realized that astrophysically realistic
black holes are expected to be surrounded by conducting plasmas
which may confine and reflect the electromagnetic (spin-$1$)
perturbations modes. Likewise, deviations of the perturbed black
hole from spherical symmetry are expected to produce gravitational
(spin-$2$) wave modes which propagate in the black-hole spacetime.

In the present paper we shall use {\it analytical} techniques in
order to explore the physical properties of these higher-spin
black-hole-mirror bombs. In particular, the aim of the present paper
is twofold:
\newline
(1) As emphasized above, it has been established \cite{CarDias} that
the frequency $\Re\omega=m\Omega_{\text{H}}$ corresponds to a
marginally stable {\it stationary} resonance (with $\Im\omega=0$) of
the {\it scalar} black-hole-mirror bomb. In the present paper we
shall point out the (seemingly unknown) fact that, for confined {\it
higher}-spin fields, the frequency $\Re\omega=m\Omega_{\text{H}}$
may correspond to {\it non}-stationary unstable modes (with
$\Im\omega>0$) of the electromagnetic and gravitational
black-hole-mirror bombs.
\newline
(2) The second goal is to prove analytically that the instability
growth rates (the values of $\Im\omega$) which characterize the
confined higher-spin fields are much larger than the instability
growth rate (\ref{Eq3}) reported for the specific case of confined
scalar (spin-$0$) fields. This fact, to be proved below, suggests
that the higher-spin (electromagnetic and gravitational)
black-hole-mirror bombs are much more explosive than the original
black-hole-mirror-scalar-field bomb studied in \cite{CarDias}.

\section{Description of the system}

The physical system we consider consists of a massless bosonic field
$\Psi$ linearly coupled to a spinning Kerr black hole of mass $M$
and angular momentum $Ma$. In the Boyer-Lindquist coordinate system
$(t,r,\theta,\phi)$ the black-hole spacetime is described by the
line-element \cite{Chan,Kerr}
\begin{eqnarray}\label{Eq4}
ds^2=-\Big(1-{{2Mr}\over{\rho^2}}\Big)dt^2-{{4Mar\sin^2\theta}\over{\rho^2}}dt
d\phi+{{\rho^2}\over{\Delta}}dr^2
%\nonumber \\
+\rho^2d\theta^2+\Big(r^2+a^2+{{2Ma^2r\sin^2\theta}\over{\rho^2}}\Big)\sin^2\theta
d\phi^2,
\end{eqnarray}
where $\Delta\equiv r^2-2Mr+a^2$ and $\rho^2\equiv
r^2+a^2\cos^2\theta$. The zeroes of $\Delta$:
\begin{equation}\label{Eq5}
r_{\pm}=M\pm(M^2-a^2)^{1/2}\  ,
\end{equation}
are the black-hole (event and inner) horizons. In order to
facilitate a fully analytical study, we shall assume that the black
hole is rapidly-rotating with $a\simeq M$. As shown in
\cite{CarDias}, the instability growth rate characterizing the
composed black-hole-mirror bomb is an increasing function of the
black-hole rotation parameter $a$. Hence, these rapidly-rotating
(near-extremal) black holes are expected to produce the largest
instability growth rates (the largest $\Im\omega$ values).

Teukolsky \cite{Teu} has shown that the dynamics of massless scalar,
electromagnetic, and gravitational perturbation fields in the
rotating Kerr black-hole spacetime are governed by the single master
equation \cite{Notes}:
\begin{eqnarray}\label{Eq6}
\Big[{{(r^2+a^2)^2}\over{\Delta}}-a^2\sin^2\theta\Big]
{{\partial^2\Psi}\over{\partial
t^2}}+{{4Mar}\over{\Delta}}{{\partial^2\Psi}\over{\partial t
\partial\phi}}+\Big({{a^2}\over{\Delta}}
-{{1}\over{\sin^2\theta}}\Big){{\partial^2\Psi}\over{\partial\phi^2}}
\nonumber \\ -\Delta^{-s}{{\partial}\over{\partial
r}}\Big(\Delta^{s+1}{{\partial\Psi}\over{\partial
r}}\Big)-{{1}\over{\sin\theta}}{{\partial}\over{\partial\theta}}\Big(\sin\theta
{{\partial\Psi}\over{\partial\theta}}\Big)-2s\Big[{{a(r-M)}\over{\Delta}}
+{{i\cos\theta}\over{\sin^2\theta}}\Big]{{\partial\Psi}\over{\partial\phi}}
\nonumber \\
-2s\Big[{{M(r^2-a^2)}\over{\Delta}}-r
-ia\cos\theta\Big]{{\partial\Psi}\over{\partial
t}}+(s^2\cot^2\theta-s)\Psi=0\  .
\end{eqnarray}

Following \cite{TeuPre2} we shall use the ``ingoing Kerr" coordinate
system $(v,r,\theta,\bar\phi)$ which is well behaved on the
black-hole event horizon. These coordinates are related to the
Boyer-Lindquist coordinates by the transformation equations
\cite{TeuPre2}
\begin{equation}\label{Eq7}
dv=dt+(r^2+a^2)dr/\Delta\ \ \ \text{and}\ \ \
d\bar\phi=d\phi+adr/\Delta\  .
\end{equation}
 Decomposing the field
$\Psi$ in the form \cite{Notedec}
\begin{equation}\label{Eq8}
{_s\Psi_{lm}}=\sum_{l,m}e^{im\bar\phi}{_sS_{lm}}(\theta;a\omega){_sR_{lm}}(r;a,\omega)e^{-i\omega
v}\ ,
\end{equation}
one finds \cite{Teu} that the radial function ${_sR_{lm}}$ and the
angular function ${_sS_{lm}}$ obey differential equations of the
confluent Heun type \cite{TeuPre2,Heun,Flam,Fiz1}, see Eqs.
(\ref{Eq9}) and (\ref{Eq11}) below.

It is worth emphasizing that the stability (or instability) of each
mode is determined by the sign of $\Im\omega$: stable (decaying in
time) modes are characterized by $\Im\omega<0$ whereas unstable
(growing in time) modes are characterized by $\Im\omega>0$.
Stationary (marginally stable) solutions are characterized by
$\Im\omega=0$.

The angular functions ${_sS_{lm}}(\theta;a\omega)$ are known as the
spin-weighted spheroidal harmonics. These functions are solutions of
the angular equation \cite{TeuPre2,Heun,Flam,Hodop}
\begin{equation}\label{Eq9}
{1\over {\sin\theta}}{\partial \over
{\partial\theta}}\Big(\sin\theta {{\partial
{_sS_{lm}}}\over{\partial\theta}}\Big)+\Big[a^2\omega^2\cos^2\theta-2a\omega
s\cos\theta-{{(m+s\cos\theta)^2}\over{\sin^2\theta}}-s^2+{_sE_{lm}}\Big]{_sS_{lm}}=0\
\end{equation}
in the interval $\theta\in [0,\pi]$. The boundary conditions of
regularity at the poles $\theta=0$ and $\theta=\pi$ pick out a
discrete set of eigenvalues ${_sE_{lm}}(a\omega)$ labeled by the
integers $m$ and $l\geq\text{max}\{|m|,|s|\}$. These angular
eigenvalues can be expanded in the form \cite{TeuPre1,BerAlm}
\begin{equation}\label{Eq10}
{_sE_{lm}}(a\omega)=l(l+1)+\sum_{k=1}^{\infty}c_k(a\omega)^k\  ,
\end{equation}
where the expansion coefficients $\{c_k(s,l,m)\}$ are given in
\cite{TeuPre1,BerAlm}.

The radial function ${_sR_{lm}}(r;a,\omega)$ satisfies the
differential equation \cite{TeuPre2}
\begin{equation}\label{Eq11}
\Delta{{d^2{_sR_{lm}}}\over{dr^2}}+2[(1+s)(r-M)-iK]{{d{_sR_{lm}}}\over{dr}}-[2(1+2s)i\omega
r+{_s\lambda_{lm}}]{_sR_{lm}}=0\  ,
\end{equation}
where $K\equiv(r^2+a^2)\omega-ma$ and
\begin{equation}\label{Eq12}
{_s\lambda_{lm}}\equiv {_sE_{lm}}-2ma\omega+a^2\omega^2-s(s+1)\  .
\end{equation}
Note that the angular equation (\ref{Eq9}) and radial equation
(\ref{Eq11}) are coupled by the angular eigenvalues
$\{{_sE_{lm}}(a\omega)\}$. (We shall henceforth omit the indexes $s,
l$ and $m$ for brevity.)

It is convenient to define new dimensionless variables
\begin{equation}\label{Eq13}
x\equiv {{r-r_+}\over {r_+}}\ \ ;\ \ \tau\equiv 8\pi
MT_{\text{BH}}={{r_+-r_-}\over {r_+}}\ \ ;\ \ k=2\omega r_+\ \ ; \ \
\varpi\equiv 4M(\omega-m\Omega_{\text{H}})\ ,
\end{equation}
in terms of which the radial wave equation (\ref{Eq11}) becomes
\begin{equation}\label{Eq14}
x(x+\tau){{d^2R}\over{dx^2}}-\{ikx^2+x[2ik-2(1+s)]-(1+s)\tau+i\varpi\}
{{dR}\over{dx}}-[\lambda+ik(1+2s)(x+1)]R=0\
.
\end{equation}

We are interested in solutions of the radial wave equation
(\ref{Eq14}) with the physical boundary condition of purely ingoing
waves (as measured by a comoving observer) crossing the black-hole
horizon \cite{TeuPre2}. As shown in \cite{TeuPre2}, this physically
motivated boundary condition corresponds to the behavior
\begin{equation}\label{Eq15}
R(x\to x_+)\to 1\
\end{equation}
of the radial function at the black-hole horizon. In addition,
following \cite{CarDias,Hodch} we shall assume that the field
vanishes at the location $x_{\text{m}}$ of the mirror \cite{Notetv}:
\begin{equation}\label{Eq16}
R(x=x_{\text{m}})=0\  .
\end{equation}
The boundary conditions (\ref{Eq15})-(\ref{Eq16}) single out a
discrete family of complex frequencies (labeled by the integer
resonance-parameter $n$). These characteristic frequencies are known
as Boxed Quasi-Normal resonances
$\{\omega^{\text{BQN}}(n;x_{\text{m}},M,a,s,l,m)\}$
\cite{CarDias,Noteqnm}.

\section{The algebraically special resonances of the black-hole-mirror system}

As we shall now show, the superradiant instability of the composed
black-hole-mirror-field system can be studied analytically in the
regime of rapidly-rotating (near-extremal, $\tau\ll1$) black holes
enclosed in reflecting cavities which are placed in the near-horizon
region $x_{\text{m}}\ll1$. In particular, we shall assume the
following inequalities \cite{Notespec}:
\begin{equation}\label{Eq17}
\text{max}(\tau,\varpi)\ll x_{\text{m}}\ll1\  .
\end{equation}

In the near-horizon region $x\ll1$ [see Eq. (\ref{Eq17})] the radial
equation (\ref{Eq14}) can be approximated by
\begin{equation}\label{Eq18}
x(x+\tau){{d^2R}\over{dx^2}}-\{x[2ik-2(1+s)]-(1+s)\tau+i\varpi\}{{dR}\over{dx}}-[\lambda+ik(1+2s)]R=0\
.
\end{equation}
The solution of equation (\ref{Eq18}) obeying the ingoing boundary
condition (\ref{Eq15}) at the black-hole horizon is given by
\cite{TeuPre2,Abram}
\begin{equation}\label{Eq19}
R(x)={_2F_1}(s+1/2-ik+i\delta,s+1/2-ik-i\delta;s+1-i\varpi/\tau;-x/\tau)\
,
\end{equation}
where $_2F_1(a,b;c;z)$ is the hypergeometric function \cite{Abram}
and
\begin{equation}\label{Eq20}
\delta^2\equiv k^2-(s+{1\over 2})^2-\lambda\  .
\end{equation}

In the $x/\tau\gg\text{max}(1,\varpi/\tau)$ regime [see Eq.
(\ref{Eq17})] one can use the large-$z$ asymptotic behavior of the
hypergeometric function $_2F_1(a,b;c;z)$ \cite{Abram} to approximate
the radial function (\ref{Eq19}) by \cite{TeuPre2,Notereg}
\begin{equation}\label{Eq21}
R(x)={{\Gamma(s+1-i\varpi/\tau)\Gamma(2i\delta)}\over{\Gamma(s+1/2-ik+i\delta)
\Gamma(1/2+ik+i\delta-i\varpi/\tau)}}
\Big({{x}\over{\tau}}\Big)^{-(s+1/2-ik-i\delta)}+(\delta\to
-\delta)\  .
\end{equation}
The notation $(\delta\to -\delta)$ means ``replace $\delta$ by
$-\delta$ in the preceding term." We shall henceforth consider the
case of real $\delta$ \cite{Notedel}. Using the expression
(\ref{Eq21}) for the radial wave function, one can express the
mirror-like boundary condition $R(x=x_{\text{m}})=0$ [see Eq.
(\ref{Eq16})] in the form
\begin{equation}\label{Eq22}
\Big({{x_{\text{m}}}\over{\tau}}\Big)^{2i\delta}{{\Gamma(2i\delta)\Gamma(s+1/2-ik-i\delta)
\Gamma(1/2+ik-i\delta-i\varpi/\tau)}\over
{\Gamma(-2i\delta)\Gamma(s+1/2-ik+i\delta)\Gamma(1/2+ik+i\delta-i\varpi/\tau)}}=-1\
.
\end{equation}
%Taking the logarithm of both sides of Eq. (\ref{Eq21}), one finds
%the resonance condition \cite{Notem1}
%\begin{equation}\label{Eq22}
%2i\delta\ln\Big({{x_{\text{m}}}\over{\tau}}\Big)+\ln\Big[{{\Gamma(2i\delta)}\over{\Gamma(-2i\delta)}}\Big]+
%\ln\Big[{{\Gamma(s+1/2-ik-i\delta)}\over{\Gamma(1/2+ik+i\delta-i\varpi/\tau)}}\Big]+
%\ln\Big[{{\Gamma(1/2+ik-i\delta-i\varpi/\tau)}\over{\Gamma(s+1/2-ik+i\delta)}}\Big]=i\pi(2n+1)\
%,
%\end{equation}
%where the resonance parameter $n$ is an integer.

As emphasized above, it has been established in \cite{CarDias} that,
for confined {\it scalar} $s=0$ fields, the resonance frequency
$\Re\omega=m\Omega_{\text{H}}$ [see Eq. (\ref{Eq2})] corresponds to
a marginally stable mode with $\Im\omega=0$. This frequency
therefore marks the boundary between stable ($\Im\omega<0$) and
unstable ($\Im\omega>0$) black-hole-mirror-scalar-field
configurations \cite{CarDias}.

However, a close inspection of the resonance condition (\ref{Eq22})
reveals the somewhat surprising fact that, for confined higher-spin
$s\neq 0$ fields, the frequency $\Re\omega=m\Omega_{\text{H}}$ may
correspond to a {\it non}-stationary mode with $\Im\omega\neq 0$. In
particular, we shall now prove that the algebraically special
\cite{Notespe} complex frequency
\begin{equation}\label{Eq23}
\omega=m\Omega_{\text{H}}+is\times{{\tau}\over{4M}}
\end{equation}
is a valid solution of the resonance equation (\ref{Eq22}) for
electromagnetic ($s=1$) and gravitational ($s=2$) fields
\cite{Notetem}. Note that the resonance (\ref{Eq23}) is
characterized by $\Im\omega>0$. This resonance frequency thus
corresponds to an {\it unstable} mode of the
Kerr-black-hole-mirror-field system.

This property of higher-spin fields can be verified directly by
substituting $\varpi=is\times\tau$ \cite{Notetem} into the resonance
condition (\ref{Eq22}) and taking the logarithm of both sides of
this equation. One then finds the characteristic equation
\cite{Notem1}
\begin{equation}\label{Eq24}
2\delta\ln\Big({{x_{\text{m}}}\over{\tau}}\Big)=i\ln\Big[{{\Gamma(2i\delta)}\over{\Gamma(-2i\delta)}}\Big]+
i\ln\Big[{{\Gamma(s+1/2-ik-i\delta)}\over{\Gamma(s+1/2+ik+i\delta)}}\Big]+
i\ln\Big[{{\Gamma(s+1/2+ik-i\delta)}\over{\Gamma(s+1/2-ik+i\delta)}}\Big]+\pi(2n+1)\
\end{equation}
for the dimensionless radii $\{x_{\text{m}}\}$ of the mirror, where
the resonance parameter $n$ is an integer.

Inspection of the resonance equation (\ref{Eq24}) reveals the
following facts: the first three terms on the r.h.s of this equation
are of the form $i[\ln(z)-\ln(\bar z)]$ and are therefore purely
real \cite{Noteabr2}. The fourth term on the r.h.s of (\ref{Eq24})
is obviously a purely real number. One therefore concludes that the
r.h.s of (\ref{Eq24}) is purely real. Hence, the l.h.s of
(\ref{Eq24}) must also be a purely real number. This implies that,
for real values of $\delta$, the solutions to the characteristic
equation (\ref{Eq24}):
\begin{equation}\label{Eq25}
x_{\text{m}}(n)=x_{\text{m}}(n=0)\times e^{\pi n/\delta}
\end{equation}
with
\begin{equation}\label{Eq26}
x_{\text{m}}(0)=\tau\times
\Big[{{\Gamma(-2i\delta)\Gamma(s+1/2+ik+i\delta)\Gamma(s+1/2-ik+i\delta)}\over
{\Gamma(2i\delta)\Gamma(s+1/2-ik-i\delta)\Gamma(s+1/2+ik-i\delta)}}\Big]^{1/2i\delta}e^{\pi/2\delta}\
,
\end{equation}
are {\it real} numbers [otherwise, the l.h.s of (\ref{Eq24}) would
not be a purely real number].

The discrete radii (\ref{Eq25})-(\ref{Eq26}) correspond to the
possible radial locations of the mirror with the algebraically
special resonance frequency (\ref{Eq23}). In Tables \ref{Table1} and
\ref{Table2} we display these characteristic mirror radii for
equatorial $l=m$ \cite{Notelm} electromagnetic ($s=1$) and
gravitational ($s=2$) modes \cite{Notedel2}. It is worth emphasizing
again that our analysis is valid in the regime $\tau\ll
x_{\text{m}}$ [see Eq. (\ref{Eq17})]. Thus, in Tables \ref{Table1}
and \ref{Table2} we present results for mirror radii with the
property $x_{\text{m}}/\tau>10$ \cite{Noten0}.

From Tables \ref{Table1} and \ref{Table2} one learns that the radii
of the mirror $\{x_{\text{m}}(n;s,m)\}$ increase with increasing
values of the overtone number $n$ [see also Eq. (\ref{Eq25})] and
decrease with increasing values of the azimuthal harmonic index $m$.
It is worth emphasizing again that these mirror radii correspond to
the algebraically special {\it unstable} resonance frequency
(\ref{Eq23}).

\begin{table}[htbp]
\centering
%\begin{tabular}{|c|c|c|c|c|c|c|}
%\hline $\ l=m\ $ & $\delta$ & \ $x_{\text{m}}(n=0)/\tau$\ & \
%$x_{\text{m}}(n=1)/\tau$\ & \ $x_{\text{m}}(n=2)/\tau$\ &
%\ $x_{\text{m}}(n=3)/\tau$\ & \ $x_{\text{m}}(n=4)/\tau$\ \\
%\hline
%\ 1\ \ &\ \ 0.492\ \ \ &\ 3.74\ \ &\ 2225.86\ \ &\ $1.33\times 10^6$\ \ &\ $7.89\times 10^8$\ \ &\ $4.70\times 10^11$\\
%\ 2\ \ &\ \ 1.336\ \ \ &\ 2.74\ \ &\ 28.80\ \ &\ 302.36\ \ &\ 3174.61\ \ &\ 33331.40 \\
%\ 3\ \ &\ \ 2.198\ \ \ &\ 2.35\ \ &\ 9.83\ \ &\ 41.03\ \ &\ 171.32\ \ &\ 715.39\\
%\ 4\ \ &\ \ 3.055\ \ \ &\ 2.18\ \ &\ 6.10\ \ &\ 17.06\ \ &\ 47.71\ \ &\ 133.43\\
%\hline
\begin{tabular}{|c|c|c|c|c|c|}
\hline $\ l=m\ $ & $\delta$ & \ $x_{\text{m}}(n=1)/\tau$\ & \
$x_{\text{m}}(n=2)/\tau$\ &
\ $x_{\text{m}}(n=3)/\tau$\ & \ $x_{\text{m}}(n=4)/\tau$\ \\
\hline
\ 1\ \ &\ \ 0.492\ \ \ &\ 2225.86\ \ &\ $1.33\times 10^6$\ \ &\ $7.89\times 10^8$\ \ &\ $4.70\times 10^{11}$\\
\ 2\ \ &\ \ 1.336\ \ \ &\ 28.80\ \ &\ 302.36\ \ &\ 3174.61\ \ &\ 33331.40 \\
\ 3\ \ &\ \ 2.198\ \ \ &\ --- \ \ &\ 41.03\ \ &\ 171.32\ \ &\ 715.39\\
\ 4\ \ &\ \ 3.055\ \ \ &\ --- \ \ &\ 17.06\ \ &\ 47.71\ \ &\ 133.43\\
\hline
\end{tabular}
\caption{The scaled mirror radii $x_{\text{m}}/\tau$ which
correspond to the equatorial $l=m$ electromagnetic $s=1$ modes with
the algebraically special resonance frequency (\ref{Eq23}). We
display results for the resonances $n=1$ to $n=4$ with the property
$x_{\text{m}}(n)/\tau>10$. Also shown are the corresponding values
of the parameter $\delta$ [see Eq. (\ref{Eq20})]. One finds that the
radius of the mirror $x_{\text{m}}(n;s=1,m)$ is an increasing
function of the overtone number $n$ and a decreasing function of the
azimuthal harmonic index $m$.} \label{Table1}
\end{table}

\begin{table}[htbp]
\centering
%\begin{tabular}{|c|c|c|c|c|c|c|}
%\hline $\ l=m\ $ & $\delta$ & \ $x_{\text{m}}(n=0)/\tau$\ & \
%$x_{\text{m}}(n=1)/\tau$\ & \ $x_{\text{m}}(n=2)/\tau$\ & \
%$x_{\text{m}}(n=3)/\tau$\ & \ $x_{\text{m}}(n=4)/\tau$\ \\
%\hline
%\ 2\ \ &\ \ 2.051\ \ \ &\ 2.49\ \ &\ 11.54\ \ &\ 53.37\ \ &\ 246.91\ \ &\ 1142.36\\
%\ 3\ \ &\ \ 2.794\ \ \ &\ 2.24\ \ &\ 6.90\ \ &\ 21.23\ \ &\ 65.36\ \ &\ 201.24\\
%\ 4\ \ &\ \ 3.565\ \ \ &\ 2.12\ \ &\ 5.11\ \ &\ 12.33\ \ &\ 29.76\ \ &\ 71.83\\
%\ 5\ \ &\ \ 4.358\ \ \ &\ 2.04\ \ &\ 4.20\ \ &\ 8.63\ \ &\ 17.75\ \ &\ 36.51\\
%\hline
\begin{tabular}{|c|c|c|c|c|c|}
\hline $\ l=m\ $ & $\delta$ & \ $x_{\text{m}}(n=1)/\tau$\ & \
$x_{\text{m}}(n=2)/\tau$\ & \
$x_{\text{m}}(n=3)/\tau$\ & \ $x_{\text{m}}(n=4)/\tau$\ \\
\hline
\ 2\ \ &\ \ 2.051\ \ \ &\ 11.54\ \ &\ 53.37\ \ &\ 246.91\ \ &\ 1142.36\\
\ 3\ \ &\ \ 2.794\ \ \ &\ --- \ \ &\ 21.23\ \ &\ 65.36\ \ &\ 201.24\\
\ 4\ \ &\ \ 3.565\ \ \ &\ --- \ \ &\ 12.33\ \ &\ 29.76\ \ &\ 71.83\\
\ 5\ \ &\ \ 4.358\ \ \ &\ --- \ \ &\ --- \ \ &\ 17.75\ \ &\ 36.51\\
\hline
\end{tabular}
\caption{The scaled mirror radii $x_{\text{m}}/\tau$ which
correspond to the equatorial $l=m$ gravitational $s=2$ modes with
the algebraically special resonance frequency (\ref{Eq23}). We
display results for the resonances $n=1$ to $n=4$ with the property
$x_{\text{m}}(n)/\tau>10$. Also shown are the corresponding values
of the parameter $\delta$ [see Eq. (\ref{Eq20})]. One finds that the
radius of the mirror $x_{\text{m}}(n;s=2,m)$ is an increasing
function of the overtone number $n$ and a decreasing function of the
azimuthal harmonic index $m$.} \label{Table2}
\end{table}

In order to check the accuracy of our analytical results, we have
solved Eq. (\ref{Eq19}) numerically with the mirror boundary
condition (\ref{Eq16}). For the fundamental gravitational mode with
$l=m=2$ we find that the exact (numerical) result agrees with our
analytical result [see Eq. (\ref{Eq26})] to within a small
correction factor of the order of $\sim1.4\tau/x_{\text{m}}\ll1$
\cite{Noterat}.

%It is not the maximal value of $\Im\omega$.

\section{Summary and discussion}

We have used analytical tools in order to explore the superradiant
instability phenomenon of rapidly-spinning Kerr black holes enclosed
inside reflecting cavities (the composed black-hole-mirror bomb).
Imposing a mirror-like boundary condition on co-rotating higher-spin
(electromagnetic and gravitational) fields in the vicinity of the
black-hole horizon, it was shown that the confined bosonic fields
grow exponentially over time.

While most former studies of the black-hole-bomb phenomenon have
considered the specific case of confined scalar ($s=0$) fields, in
the present study we have focused on the physically interesting
cases of confined electromagnetic ($s=1$) and gravitational ($s=2$)
fields.

Former studies \cite{CarDias} of the black-hole-mirror-scalar-field
system have established that the critical frequency
$\Re\omega=m\Omega_{\text{H}}$ [see Eq. (\ref{Eq2})] marks the onset
of instability in this composed system \cite{Notesca} (that is,
$\Re\omega=m\Omega_{\text{H}}$ corresponds to a {\it stationary}
scalar mode with $\Im\omega=0$). On the other hand, in the present
study we have revealed the (somewhat surprising) fact that this
property is actually {\it not} a generic feature of composed
black-hole-mirror bombs. In particular, we have shown that, for
confined higher-spin $s\neq 0$ fields, the frequency
$\Re\omega=m\Omega_{\text{H}}$ may correspond to a {\it
non}-stationary mode with $\Im\omega\neq 0$.

Although the equations governing the dynamics of the confined fields
in the black-hole spacetime are rather complex [see, in particular,
the coupled equations (\ref{Eq9}), (\ref{Eq12}), and (\ref{Eq14})],
we have succeeded to prove analytically that the algebraically
special \cite{Notespe} frequency $\omega=m\Omega_{\text{H}}+s\cdot
i2\pi T_{\text{BH}}$ [see Eq. (\ref{Eq23})] is a valid solution of
the resonance condition (\ref{Eq22}). That is, we have shown that
this complex frequency corresponds to an unstable resonance of the
composed black-hole-mirror-field system \cite{Noteunn}.

It is worth emphasizing that our analysis is valid in the regime of
rapidly-rotating (near-extremal) black holes with $\tau\ll
x_{\text{m}}\ll1$ [see Eq. (\ref{Eq17})]. Specifically, since each
inequality sign in (\ref{Eq17}) roughly corresponds to an
order-of-magnitude difference between two variables (that is,
$\tau/x_{\text{m}}\lesssim 10^{-1}$ with $x_{\text{m}}\lesssim
10^{-1}$), our results should be valid in the regime $\tau\lesssim
10^{-3}$ \cite{Notecon}. Thus, the algebraically special resonance
(\ref{Eq23}) for confined electromagnetic and gravitational fields
is characterized by
\begin{equation}\label{Eq27}
\Im\omega_{\text{max}}=O(10^{-3}M^{-1})\ .
\end{equation}

The present analysis provides compelling evidence that the
higher-spin (electromagnetic and gravitational) black-hole-mirror
bombs are much more explosive than the original
black-hole-mirror-scalar-field bomb studied in \cite{CarDias}. In
particular, taking cognizance of Eqs. (\ref{Eq3}) and (\ref{Eq27}),
one concludes that the instability growth rates (the values of
$\Im\omega$) which characterize the superradiant confined
higher-spin $s\neq 0$ fields are much larger than the maximal growth
rate (\ref{Eq3}) which characterizes a superradiant confined scalar
$s=0$ field \cite{Notecom}.

Finally, it is worth emphasizing that the present {\it analytical}
exploration of the black-hole-bomb phenomenon is restricted to the
regime of linear perturbation fields. As we have seen, the initial
exponential growth of the confined fields (the superradiant
instability) manifests itself at this linear level. However, it is
clear that a fully non-linear numerical simulation of the coupled
black-hole-field dynamics is required in order to reveal the
end-state of this explosive superradiant instability.

\bigskip
\noindent
{\bf ACKNOWLEDGMENTS}
\bigskip

This research is supported by the Carmel Science Foundation. I thank
Yael Oren, Arbel M. Ongo and Ayelet B. Lata for helpful discussions.

%\newpage


\begin{thebibliography}{99}
\bibitem{Noteun} We use natural units in which $G=c=\hbar=1$.
\bibitem{Notemax1} An extremal (maximally spinning) Kerr black hole is characterized by
$J=M^2=A/8\pi$.
\bibitem{NoteRN} An analogous extraction of Coulomb energy may occur when
charged scalar fields interact with charged Reissner-Nordstr\"om
black holes \cite{Bekch,Hodch}.
\bibitem{Noteher} Here $\omega$ and $m$ are respectively the conserved frequency and
azimuthal harmonic index of the scalar wave field.
\bibitem{Noteimw} The instability growth rate of the field is
characterized by the imaginary part $\Im\omega$ of its frequency.
\bibitem{Notemax} Note that this (maximal) value of $\Im\omega^{\text{scalar}}$ corresponds to an
instability timescale $T^{\text{scalar}}_{\text{ins}}\equiv
1/\Im\omega^{\text{scalar}}\simeq {1\over 6}\times 10^5 M$ which is
much longer than the dynamical timescale set by the mass $M$ of the
black hole.
\bibitem{Notes} The dependent variables in the black-hole perturbation equations are
related to null tetrad components of the electromagnetic and
gravitational (Weyl) fields, see \cite{Teu} for details. The
parameter $s$ in the Teukolsky master equation (\ref{Eq6}) is known
as the spin-weight of the field. This parameter takes different
values for different perturbation fields \cite{Teu}. In particular,
a scalar perturbation field $\phi$ is characterized by $s=0$; the
electromagnetic perturbation field $\phi_0$ is characterized by
$s=1$ and the electromagnetic perturbation field
$(r-ia\cos\theta)^2\phi_2$ is characterized by $s=-1$; the
gravitational perturbation field $\psi_0$ is characterized by $s=2$
and the gravitational perturbation field $(r-ia\cos\theta)^4\psi_4$
is characterized by $s=-2$. (see \cite{Teu} for the precise
definitions of the variables $\phi_0, \phi_2, \psi_0$, and
$\psi_4$).
\bibitem{Notedec} Here $l$ and $m$ are respectively the spheroidal and azimuthal
harmonic indexes of the mode [see Eq. (\ref{Eq9}) below], and
$\omega$ is the conserved frequency of the wave field.
\bibitem{Notetv} Note that in the Boyer-Lindquist coordinate system one should replace
the expansion (\ref{Eq8}) by
${_s\Psi_{lm}}=\sum_{l,m}e^{im\phi}{_sS_{lm}}(\theta;a\omega){_sR^{\text{BL}}_{lm}}(r;a,\omega)e^{-i\omega
t}$. The relation between the radial functions $R$ and
$R^{\text{BL}}$ is given by
$R^{\text{BL}}=R\cdot\exp[-i\int(K/\Delta)dr]$ [Here we have used
the relations $v=t+\int[(r^2+a^2)/\Delta]dr$ and
$\bar\phi=\phi+\int(a/\Delta)dr$, see Eq. (\ref{Eq7})]. Thus, the
functions $R$ and $R^{\text{BL}}$ share the same set of radial
zeroes.
\bibitem{Noteqnm} Note that the extensively studied family of black-hole quasinormal resonances is
characterized by the boundary condition of outgoing waves at spatial
infinity \cite{TeuPre2,TeuPre1}.
\bibitem{Notespec} Specifically, below we shall assume the inequalities
$x_{\text{m}}<10^{-1}$ and $\tau/x_{\text{m}}<10^{-1}$.
\bibitem{Notereg} It is worth emphasizing that the expression (\ref{Eq21}) for
the radial wave function $R(x)$ is valid in the regime $\tau\ll
x\ll1$, see Eq. (\ref{Eq17}).
\bibitem{Notedel} One can choose $\delta>0$ without loss of generality.
\bibitem{Notetem} Note that the resonance frequency (\ref{Eq23}) corresponds
to $\varpi=is\times\tau$ [see Eq. (\ref{Eq13})]. In addition, note
that the imaginary part of this unstable resonance can be expressed
in terms of the black-hole temperature as $\Im\omega=s\cdot 2\pi
T_{\text{BH}}$.
\bibitem{Notem1} Here we have used the relation $-1=e^{i\pi(2n+1)}$ which implies $\ln(-1)=i\pi(2n+1)$,
where $n$ is an integer.
\bibitem{Notelm} It is well known that these equatorial $l=m$ modes are
the most amplified ones \cite{CarDias,TeuPre2}.
\bibitem{Notedel2} We are interested here in the relation
$x_{\text{m}}=x_{\text{m}}(\tau)$ to leading-order in the small
parameter $\tau\ll1$. We therefore take $a\omega=m/2+O(\tau)$ and
$k=m+O(\tau^2)$ in Eqs. (\ref{Eq10}), (\ref{Eq12}), (\ref{Eq20}),
(\ref{Eq25}), and (\ref{Eq26}). The leading-order values of the
constants ${_s\delta_{lm}}$ are given in Tables \ref{Table1} and
\ref{Table2}.
\bibitem{Noten0} We find that modes with $n\leq 0$ are characterized by
$x_{\text{m}}/\tau<10$. Hence, these modes are beyond the regime of
validity of our analysis.
\bibitem{Notesca} As explained above, for the specific case of confined scalar $s=0$
fields, one finds \cite{CarDias} that scalar modes with
$\Re\omega>m\Omega_{\text{H}}$ are stable ($\Im\omega<0$) whereas
scalar modes with $\Re\omega<m\Omega_{\text{H}}$ are unstable
($\Im\omega>0$). The critical scalar mode with
$\Re\omega=m\Omega_{\text{H}}$ is marginally stable (stationary)
with $\Im\omega=0$.
\bibitem{Noteunn} The fact that, for confined higher-spin $s\neq 0$ fields, the
mode with $\Re\omega=m\Omega_{\text{H}}$ corresponds to an {\it
unstable} resonance with $\Im\omega>0$, implies that the spectrum of
unstable higher-spin resonances contains frequencies with
$\Re\omega>m\Omega_{\text{H}}$. This is to be contrasted with the
scalar $s=0$ field case, in which the unstable resonances are
characterized by $\Re\omega<m\Omega_{\text{H}}$.
\bibitem{Notecon} It should be realized that here we have used a very
conservative estimate for the regime of validity of our analysis. In
fact, the two inequalities $\tau\ll x_{\text{m}}\ll1$
($\tau/x_{\text{m}}\lesssim 10^{-1}$ with $x_{\text{m}}\lesssim
10^{-1}$) may be satisfied simultaneously in the less restrictive
regime $\tau\lesssim 10^{-2}$.
\bibitem{Notecom} In particular, equations (\ref{Eq3}) and (\ref{Eq27}) imply
that the electromagnetic and gravitational black-hole-mirror bombs
are characterized by instability timescales which are at least two
orders of magnitude shorter than the instability timescale of the
scalar black-hole-mirror bomb.
\end{thebibliography}
\end{document}